\documentstyle[emulateapj,pstricks,psfig]{article}

\newcommand{\beqa}{\begin{eqnarray}}
\newcommand{\eeqa}{\end{eqnarray}}

\newcommand{\hkpc}{\ h^{-1}{\rm kpc}}
\newcommand{\hMsun}{\ h^{-1}M_{\odot}}

\newcommand{\gsim}{\lower .1ex\hbox{\rlap{\raise .6ex\hbox{\hskip .3ex
        {\ifmmode{\scriptscriptstyle >}\else
                {$\scriptscriptstyle >$}\fi}}}
        \kern -.4ex{\ifmmode{\scriptscriptstyle \sim}\else
                {$\scriptscriptstyle\sim$}\fi}}}
\newcommand{\lsim}{\lower .1ex\hbox{\rlap{\raise .6ex\hbox{\hskip .3ex
        {\ifmmode{\scriptscriptstyle <}\else
                {$\scriptscriptstyle <$}\fi}}}
        \kern -.4ex{\ifmmode{\scriptscriptstyle \sim}\else
                {$\scriptscriptstyle\sim$}\fi}}}
\newcommand{\beq}{\begin{equation}}
\newcommand{\eeq}{\end{equation}}
\newcommand{\ks}{\rm ~km~s^{-1}}

\newcommand{\Rvir}{R_{\rm vir}}
\newcommand{\Mvir}{M_{\rm vir}}

\newcommand{\zre}{z_{\rm re}}

\begin{document}
\slugcomment{{\em Astrophysical Journal, submitted}}
\lefthead{REIONIZATION AND THE ABUNDANCE OF GALACTIC SATELLITES}
\righthead{BULLOCK, KRAVTSOV, \& WEINBERG}

\title{REIONIZATION AND THE ABUNDANCE OF GALACTIC SATELLITES}\vspace{3mm}

\author{James S. Bullock\altaffilmark{1}, Andrey V. Kravtsov\altaffilmark{2} 
and David H. Weinberg}
\affil{Department of Astronomy, The Ohio State University,
    140 W. 18th Ave, Columbus, OH 43210-1173}

\altaffiltext{1}{james,andrey,dhw@astronomy.ohio-state.edu}
\altaffiltext{2}{Hubble Fellow}

\begin{abstract}
One  of  the  main challenges facing   standard hierarchical structure
formation models is that the  predicted abundance of galactic subhalos
with circular velocities $v_c \sim  10-30\ks$ is an order of magnitude
higher  than  the number of  satellites actually  observed  within the
Local Group.  Using a simple model  for the formation and evolution of
dark halos, based on the extended Press-Schechter formalism and tested
against  N-body results, we show that  the theoretical predictions can
be reconciled with observations if  gas accretion in low-mass halos is
suppressed after   the epoch of  reionization.   In this  picture, the
observed dwarf  satellites correspond to  the small  fraction of halos
that accreted  substantial amounts of    gas before reionization.  The
photoionization mechanism   naturally  explains why   the  discrepancy
between predicted  halos and observed satellites sets  in at $v_c \sim
30\ks$, and for reasonable choices of the reionization redshift ($\zre
\sim  5-12$) the model can reproduce  both the  amplitude and shape of
the observed   velocity  function  of galactic satellites.     If this
explanation is correct, then typical  bright galaxy halos contain many
low-mass dark matter  subhalos.  These   might be detectable   through
their  gravitational   lensing effects,   through their   influence on
stellar  disks, or  as dwarf satellites   with very high mass-to-light
ratios.  This model also predicts a diffuse stellar component produced
by large  numbers of tidally  disrupted dwarfs, perhaps  sufficient to
account for most of the Milky Way's stellar halo.
\end{abstract}
\keywords{cosmology: theory -- galaxies:formation}


\section{Introduction}


Cold Dark Matter (CDM) models with scale-invariant initial fluctuation
spectra define  an elegant and well-motivated  class  of theories with
marked  success at   explaining most   properties  of   the local  and
high-redshift universe.  One   of the few  perceived  problems of such
models is the apparent overprediction of the number of satellites with
circular velocities $v_c \sim   10-30 \ks$ within the  virialized dark
halos of the Milky Way and M31 (Klypin et al. 1999a, hereafter KKVP99;
Moore et al. 1999; see also Kauffmann et  al. 1993 and Gonzales et al.
1998).   The  most  recent of    these investigations   are based   on
dissipationless simulations,  so  the problem may  be rephrased  as  a
mismatch between  the expected number of  {\em dark matter\/} subhalos
orbiting within  the Local Group and  the observed number of satellite
galaxies.  In order to overcome  this difficulty, several authors have
suggested modifications to the standard  CDM scenario.  These  include
(1)   reducing   the small-scale power    by  either  appealing   to a
specialized   model of inflation  with broken  scale invariance (e.g.,
Kamionkowski \& Liddle 1999) or substituting  Warm Dark Matter for CDM
(e.g., Hogan 1999), and (2) allowing for strong self-interaction among
dark matter  particles, thereby enhancing satellite destruction within
galactic  halos  (e.g., Spergel and  Steinhardt   1999;  but  see  the
counter-argument by Miralda-Escud\'e 2000).  In this paper, we explore
a more conservative solution.

Many authors have  pointed out that  accretion of  gas onto low-mass
halos and subsequent star formation are inefficient in the presence of
a strong  photoionizing background (Ikeuchi  1986; Rees 1986, Babul \&
Rees 1992; Efstathiou 1992; Shapiro,  Giroux, \& Babul 1994; Thoul  \&
Weinberg  1996; Quinn, Katz, \&  Efstathiou 1996).  Motivated by these
results, we  investigate whether  the abundance of  satellite galaxies
can be explained in hierarchical  models if low-mass galaxy  formation
is suppressed after  reionization, and we  propose that the observable
satellites  correspond to  those   halos that  accreted a  substantial
amount of gas before reionization.   This solution is similar in  some
ways  to the  idea that supernova   feedback ejects  gas from low-mass
halos (Dekel \&  Silk 1986), but it offers  a natural explanation  for
why  some satellite  galaxies at each circular velocity
 survive,  while most  are too dim  to be
seen.   In addition,    the  reionization  solution  seems  physically
inevitable, while the feedback mechanism  may be inadequate in all but
the smallest halos (Mac Low \& Ferrara 1998).

Our approach to calculating the  satellite abundance uses an extension
of the Press-Schecter (1974)  formalism to predict the  mass accretion
history of galactic halos and a simple  model for orbital evolution of
the  accreted  substructure.  For our  analysis,  we adopt  a flat CDM
model  with  a non-zero  vacuum energy  and the  following parameters:
$\Omega_m = 0.3,
\Omega_{\Lambda} = 0.7, h=0.7, \sigma_8=1.0$, where $\sigma_8$ is the
rms fluctuation  on the  scale of  $8h^{-1}$ Mpc,   $h$ is the  Hubble
constant in    units  of $100  \ks {\rm    Mpc}^{-1}$,  and $\Omega_m$  and
$\Omega_{\Lambda}$ are the density  contributions  of matter and   the
vacuum respectively, in units of the critical density.

{\pspicture(0.5,4.9)(12.0,18.) 
\rput[tl]{0}(0.2,18.4){\epsfxsize=9.cm 
\epsffile{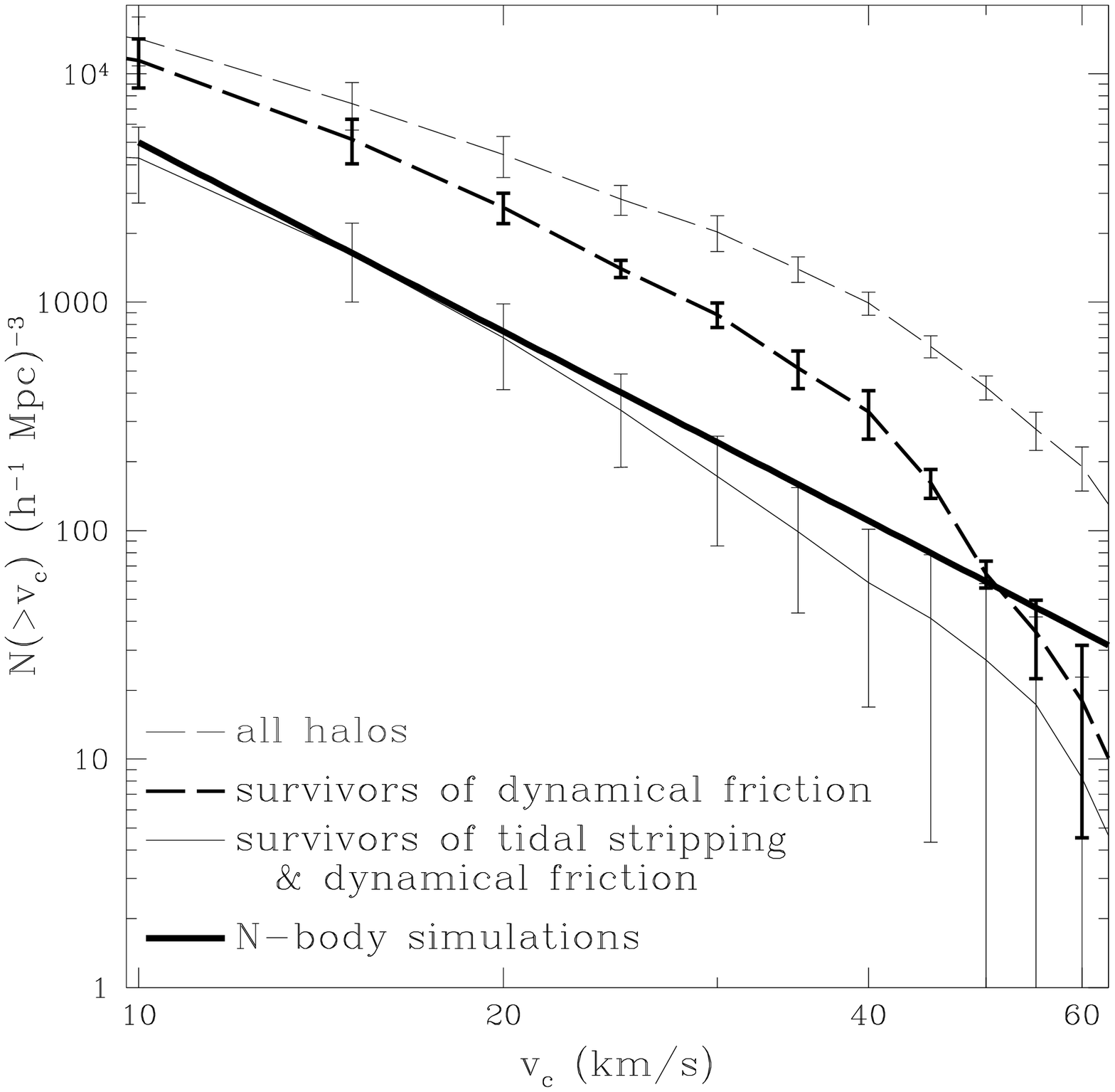}} 
\rput[tl]{0}(.1,9){ 
\begin{minipage}{8.7cm} 
  \small\parindent=4.5mm {\sc Fig.}~1.--- Average cumulative velocity
  function of dark matter halos accreted by halos of mass $1.1\times
  10^{12}h^{-1}{\ }{\rm M_{\odot}}$ ($z=0$). The average is over 300
  merger histories.  The {\em thin dashed line}
  represents velocity function of all accreted halos; the {\em thick 
    dashed line} corresponds to halos that survive effects of dynamical
    friction; the {\em thin solid line} corresponds to halos that survive
    both the dynamical friction and tidal disruption. The {\em thick
      solid line} is a fit to results of cosmological simulations
    presented in KKVP99.  The figure shows that a substantial fraction
    of accreted substructures can be destroyed by $z=0$; most of the
    destruction occurs relatively early ($z\gsim 0.5$).
\end{minipage} 
}
\endpspicture}

\section{Method}

\subsection{Modeling the Subhalo Distribution}

Because  N-body   simulations   that resolve  low-mass    subhalos are
computationally expensive, we  develop  an approximate analytic  model
for the  accretion  history and  orbital evolution of  satellite halos
within a typical Milky Way-size dark halo. Our general procedure is to
use the extended  Press-Schechter method (Bond  et al.  1991; Lacey \&
Cole 1993, hereafter LC93) to  construct the subhalo accretion history
for each  galactic halo and then to  determine which accreted subhalos
are dragged to  the  center due to dynamical  friction  and  which are
tidally destroyed.   The subhalos  that  survive at  $z=0$ are used to
construct the final velocity function of satellite halos.

We assume that  the density profile of each  halo is  described by the
NFW form (Navarro, Frenk, \&  White 1997): $\rho_{\rm NFW}(x)  \propto
x^{-1}(1+x)^{-2}$, where  $x=r/r_s$,  and $r_{s}$  is a characteristic
inner radius.  Given a halo of mass $\Mvir$ at redshift $z$, the model
of Bullock    et al. (1999)   supplies  the  typical $r_s$   value and
specifies  the   profile completely.    The circular   velocity curve,
$v^2(r) \equiv GM(r)/r$, peaks at the radius  $r_{\rm max} \simeq 2.16
r_{\rm s}$.  Throughout this paper, we use $v_c$ to refer to the 
circular velocity at $r_{\rm max}$.

We begin by constructing mass growth  and halo accretion histories for
an ensemble of galaxy-sized dark matter halos.  This is done using the
merger tree method of Somerville \& Kolatt (1999, hereafter SK99).  We
modified the method  slightly  to require that  at each  time step the
number of progenitors in  the mass range of  interest be close to  the
expected  average.   This    modification significantly  improves  the
agreement  between   the generated  progenitor mass   function and the
analytic prediction.  We start with halos of mass $M_{\rm vir}=1.1
\times 10^{12} \hMsun$, at $z=0$,  corresponding  to $v_c = 220  \ks$,
and trace satellite accretion   histories  back to $z=10$ using   time
steps chosen as described in SK98.   We track only accreted halos that
are more massive  than $M_m = 7 \times  10^6 h^{-1} M_{\odot}$,  which
corresponds to $v_c\simeq 10
\ks$  at   $z=10$ and $v_c   \simeq 4  \ks$  at $z=0$,   and treat the
accretion of halos smaller than $M_m$ as diffuse mass growth.

For each  step back   in  time, the merger   tree  provides a list  of
progenitors.  The most  massive  progenitor  is identified  with   the
galactic halo and the  rest of the   progenitors (with $M >  M_m$) are
recorded  as accreted substructure.  We are  left with a record of the
mass growth for  the galactic halo,  $M_{\rm vir}(z)$, as well as  the
mass  of each  accreted  subhalo,  $M_a$,  and  the  redshift  of  its
accretion,  $z_a$.  For  the   results  presented below,   we  use 300
ensembles of formation histories  for galactic host halos. Our results
do not change if the number of ensembles is increased.

Each subhalo  is assigned  an initial orbital
circularity, $\epsilon$, defined as the  ratio of the angular momentum
of the subhalo   to that of  a  circular orbit  with the  same  energy
$\epsilon  \equiv J/J_{c}$.  We choose $\epsilon$  by drawing a random
number uniformly distributed from 0.1  to 1.0, an approximation to the
distribution  found in cosmological simulations  (Ghigna et al. 1998).
Our  results are not sensitive to  the choice of the  above range. 
To determine whether the accreted halo's orbit will decay,
we use  Chandrasekhar's   formula    to  calculate  the    decay    time,
$\tau_{DF}^c$, of the orbit's  circular  radius $R_c$, as  outlined in
Klypin et al. (1999b).  The circular radius is defined as the radius of
a circular orbit  with  the same energy   as the actual orbit.    Each
subhalo  is assumed  to  start   at $R_c  =   0.5 \Rvir(z_a)$,   where
$\Rvir(z_a)$ is the virial  radius  of the host halo   at the time  of
accretion.  This  choice   approximately  represents typical   binding
energies of subhalos in simulations.   
The amplitude of the resulting velocity function
is somewhat  sensitive  to  this  choice.   If we   adopt $R_c =  0.75
\Rvir(z_a)$ (less  bound  orbits),  the amplitude  increases  by $\sim
50\%$.   A change of this magnitude  will not affect our conclusions,
discussed in   \S 4.  Once $\tau_{\small  DF}^{\rm   c}$ is known, the
fitting formula of Colpi et al.  (1999) provides the appropriate decay
time for the given circularity
\beq 
\tau_{\small DF} = \tau_{\small DF}^{\rm c} \epsilon^{0.4}.  
\eeq
If $\tau_{DF}$ is smaller than the time  left between $z_a$ and $z=0$,
$\tau_{DF} \le t_0 - t_a$, then the subhalo will not survive until the
present,   and it is   therefore  removed from the   list  of galactic
subhalos.

If  $\tau_{DF}$ is too  long for the  orbit to have decayed completely
($\tau_{DF} > t_0  - t_a$), we  check whether  the subhalo would  have
been tidally disrupted.   We assume that the  halo is disrupted if the
tidal   radius becomes smaller than  $r_{\rm  max}$.  Such a situation
means either   a  drastic reduction in   the   measured $v_c$ (quickly
pushing $v_c$ below our  range of interest) or  that the halo (and any
central baryonic   component) should   become  unstable and    rapidly
dissolve (Moore, Katz \& Lake 1996).

The tidal radius, $r_t$, is determined at the pericenter of the orbit
at $z=0$, where  the tides are the strongest. 
For an orbit that has decayed to  a final circular radius $R_c^0$, the
pericenter at this time, $R_{p}$, is given by the smaller of the roots
of the bound orbit  equation (e.g., van den Bosch et al.  1999) 
\beq
\left(\frac{R_c^0}{r}\right)^2   -  \frac{1}{\epsilon^2}  =    \frac{2
  [\Phi(R_c^0)  -   \Phi(r)]}{\epsilon^2  v^2(R_c^0)}.   
\eeq    
Here,
$\Phi(r) = - 4.6 v_c^2 \ln(1+x)/x$ is the potential  of the host halo. 
Once $R_p$ is known,  we determine $r_t$ as outlined  in Klypin et al. 
(1999b).  If $r_t \le  r_{\rm max}$ we  declare the subhalo to  be tidally
destroyed and remove it from the list of galactic subhalos.
Accreted halos  that survive ($\tau_{DF} > t_0  - t_a$ and 
$r_t > r_{\rm max}$) are assumed   to preserve  the $v_c$   
they had  when they  were accreted.

The  resulting velocity  function  of  surviving  subhalos within  the
virial radius of  the host ($\sim  200 \hkpc$ at $z=0$), averaged over
all merger histories, is shown  by the  thin solid  line in Figure  1.
The error bars reflect  the measured dispersion among different merger
histories.   The  thick  straight   line  is  the best    fit to   the
corresponding velocity  function  measured in  the cosmological N-body
simulation   of  KKVP99.  The upper  dashed    lines show the velocity
function if the effects of  dynamical friction and/or tidal disruption
are   ignored.  The   analytic  model reproduces   the  N-body results
remarkably  well.  We have  not  tuned any  parameters to  obtain this
agreement,  although  we noted  above  that  plausible  changes in the
assumed  initial circular radius    $R_c$  could change   the analytic
prediction   by  $\sim 50\%$.  The good    agreement suggests that our
analytic  model captures the   essential physics underlying the N-body
results.  An  interesting feature of  the model  is that the  subhalos
surviving at $z=0$ are  only a small  fraction  of the halos  actually
accreted, most of which are destroyed by tidal disruption.  We discuss
implications of this satellite destruction in \S4.

\subsection{Modeling Observable Satellites}

The second step in our  model is to  determine which  of the surviving
halos at $z=0$  will host observable  satellite galaxies.
The key assumption is that after the reionization redshift $\zre$, 
gas accretion is suppressed
in halos with $v_c < v_{T}$.
 We adopt a threshold  of $v_{T} = 30 \ks$,
based on the  results of  Thoul \&  Weinberg (1996),  who showed  that
galaxy formation is  suppressed in  the  presence of a   photoionizing
background for objects smaller than $\sim 30  \ks$.  This threshold was
shown to be insensitive to the assumed spectral index and amplitude of
the ionizing background (a similar result was found by Quinn et al. 1996).
Shapiro et al.\ (1997; see also Shapiro \& Raga 2000) and
Barkana \& Loeb (2000) have suggested that very low mass 
systems ($v_c \la 10\ks$)
could lose the gas they have already accreted after
reionization occurs, but we do not consider this possibility here.

The calculation of the halo velocity function in \S 2.1 is approximate
but straightforward, and we have checked its validity by comparing to 
N-body simulations.  Determining which of these halos are luminous
enough to represent known dwarf satellites requires more uncertain
assumptions about gas cooling and star formation.  We adopt a simple
model that has two free parameters: the reionization redshift $\zre$
and the fraction $f=M(\zre)/M_a$ of a subhalo's mass that must
be in place by $\zre$ in order for the halo to host an observable galaxy.

The value of $\zre$ is constrained to $\zre \ga 5$ 
by observations of high-$z$ quasars (e.g., Songaila et al. 1999)
and to $\zre \la 50$ by measurements of small-angle CMB 
anisotropies, assuming typical ranges for the cosmological parameters
(e.g., Griffiths, Barbosa, \& Liddle 1999).
The value of $f$ is constrained by the requirement that observable
halos have mass-to-light ratios in the range of observed dwarf satellites.
For the subset of (dwarf irregular) satellite galaxies with well-determined
masses, the mass-to-light ratios span the range 
$M/L_V \simeq 5 - 30$ (Mateo 1998); dwarf spheroidals have
similar $M/L_V$, but with a broader range and larger observational
uncertainties.  We can estimate $M/L_V$ for model galaxies by
assuming that they accrete a baryon mass $fM_a(\Omega_b/\Omega_m)$
before $\zre$ and convert that accreted gas with efficiency
$\epsilon_*$ into a stellar population with mass-to-light
ratio $M_*/L_V$, obtaining
\beq
 \left(\frac{M}{L_V}\right) = f^{-1} \left(\frac{\Omega_m}{\Omega_b}\right)
    \left(\frac{M_*}{L_V}\right) \epsilon_*^{-1} F_o.
\eeq
The factor $F_o$ is the fraction of the halo's virial mass $M_a$
that lies within its final optical radius (which may itself be
affected by tidal truncation).
For the $M/L_V$ values quoted above, the optical radius
is typically $\sim   2 \rm{kpc}$  (Mateo 1998),  and
representative mass profiles of surviving halos imply $F_o(2 \rm{kpc})
\simeq 0.5$; however, this factor must be considered uncertain
at the factor of two level.  The value of $\epsilon_*$ is also
uncertain because of the uncertain influence of supernova feedback,
but by definition $\epsilon_* \leq 1$.
Adopting a value $M_*/L_V \simeq 0.7$ typical for galactic
disk stars (Binney \& Merrifield 1998), 
$\Omega_m/\Omega_b \simeq 7$ (based on $\Omega_m=0.3,$ $h=0.7$,
and $\Omega_b h^2 \simeq 0.02$ from Burles \& Tytler [1998]),
$\epsilon_* = 0.5$, and $F_o=0.5$, we obtain $(M/L_V) \simeq 5 f^{-1}$.
Matching the mass-to-light ratios of typical dwarf satellite
galaxies then implies $f \sim 0.3$.  With the uncertainties
described above, a range $f \sim 0.1-0.8$ is plausible, and
the range in observed $(M/L_V)$ could reflect in large part
the variations in $f$ from galaxy to galaxy.
Values of $f \la 0.1$ would imply excessive mass-to-light ratios, 
unless the factor $F_o$ can be much smaller than we have assumed.

In sum, the two parameters that determine the fraction of
surviving halos that are observable are $\zre$ and $f$,
with plausible values in the range $\zre \sim 5-50$ and $f \sim 0.1-0.8$.
For a given subhalo of 
mass  $M_a$ and accretion redshift $z_a$,  we use equation (2.26) 
of LC93  to
probabilistically  determine the redshift $z_f$  when the main
progenitor of the  subhalo was first more massive  than $M_f = f M_a$.
We associate the subhalo  with  an observable galactic  satellite  only if
$z_f \ge \zre$.

\section{Results}

Figure 2 shows results of our model for the  specific choices of $\zre
= 8$ and $f=0.3$.  The thin solid line is the velocity function of all
surviving subhalos at $z=0$, reproduced from Figure 1.  The thick line
and shaded region shows the average and scatter in the expected number
of observable satellites ($z_f > \zre$).  The solid triangles show the
observed satellite galaxies  of the Milky  Way and M31 within radii of
$200 \hkpc$ from the centers of each galaxy (note that all results are
scaled to a fiducial  volume $1 h^{-3} \rm{Mpc}^{3}$).   We see  that the
theoretical velocity function  for  visible  satellites  is consistent
with that observed, and that the total number of observable systems is
$\sim 10\%$  of the total dark  halo  abundance.  The reason  for this
difference  is that most  of the  halos form  after reionization.  The
observable  satellites      are within    halos   that  formed  early,
corresponding to rare, high peaks in the initial density field.

Other combinations of  $\zre$    and $f$ can provide similar results
 because the fraction  of mass in place by
the epoch of reionization is larger if  reionization occurs later.  We
find that the  following pairs of choices  also reproduce the observed
galactic    satellite  velocity   function:   
$(f,\zre)=(0.1,12)$;$(0.2,10)$;$(0.4,7)$;$(0.5,5)$.
   It  is  encouraging  that successful
parameter  choices fall  in the range  of  naturally expected  values.
However, our model fails if reionization occurs too early ($\zre \gsim
12$),  since the value   of  $f$ required  to  reproduce  the observed
velocity function  would imply excessively large  $M/L_V$  ratios.  We
conclude   that the observed abundance of   satellite  galaxies may be
explained in the CDM scenario if galaxy formation in low-mass halos is
suppressed after the epoch of reionization, provided that $\zre \lsim 12$.

\section{Discussion}

The suppression of gas accretion by the photoionizing background
offers an attractive solution to the dwarf satellite problem.
The physical mechanism seems natural, almost inevitable, and requires
no fine tuning of the primordial fluctuation spectrum or properties
of the dark matter.  Although, in principle, a similar explanation
could be obtained with supernova feedback, reionization
naturally explains why most subhalos are dark, 
while the fraction $\sim 10\%$ that accreted a substantial fraction of 
their mass before $\zre$ remain visible today.
With feedback, it is not obvious why any dwarf satellites would
survive, and certainly not the specific number observed.
Photoionization also naturally explains why the discrepancy with CDM 
predictions appears at $v_c \sim 30 \ks$, rather than at a higher or
lower $v_c$.  In contrast, studies of the feedback mechanism
suggest that it is difficult to achieve the required suppression of star
formation in halos with $v_c \gsim 15 \ks$ (Mac Low \& Ferrara 1998).

In order to obtain a reasonable $M/L$ ratio  for satellite galaxies in
this scenario, the reionization redshift must be relatively low, $\zre
\lsim 12$.   For higher $\zre$,  it  is hard to understand  how  dwarf
galaxies with $v_c < 30  \ks$ could have formed  at all, though a blue
power spectrum or  non-Gaussian primordial fluctuations could help, or
perhaps    dwarfs    could  form  by    fragmentation   within  larger
proto-galaxies.  Currently, the  reionization redshift is  constrained
only within the rather broad range  $\zre \sim 5-50$,  but it might be
determined    in the future by  CMB   experiments  or by spectroscopic
studies of luminous high-$z$ objects.

A clear prediction   that  distinguishes this model  from  models with
suppressed small-scale power  or self-interacting dark matter is  that
there  should be a large number   of low-mass subhalos associated with
the  Milky Way  and similar  galaxies.   If  we assume  that the model
presented in  \S 2.2 applies  in all  cases, then  the observed  dwarf
satellites  should  be  just the  low  $M/L$ tail    of the underlying
population.   For example, in our fiducial   case (Figure 2, $\zre=8$,
$f=0.3$), reducing $f$ by a factor of 3 --- corresponding to an
\textit{increase} in the average $M/L$ by a 
{\pspicture(0,0)(2,13.) 
\rput[tl]{0}(0,13.){\epsfxsize=9.cm 
\epsffile{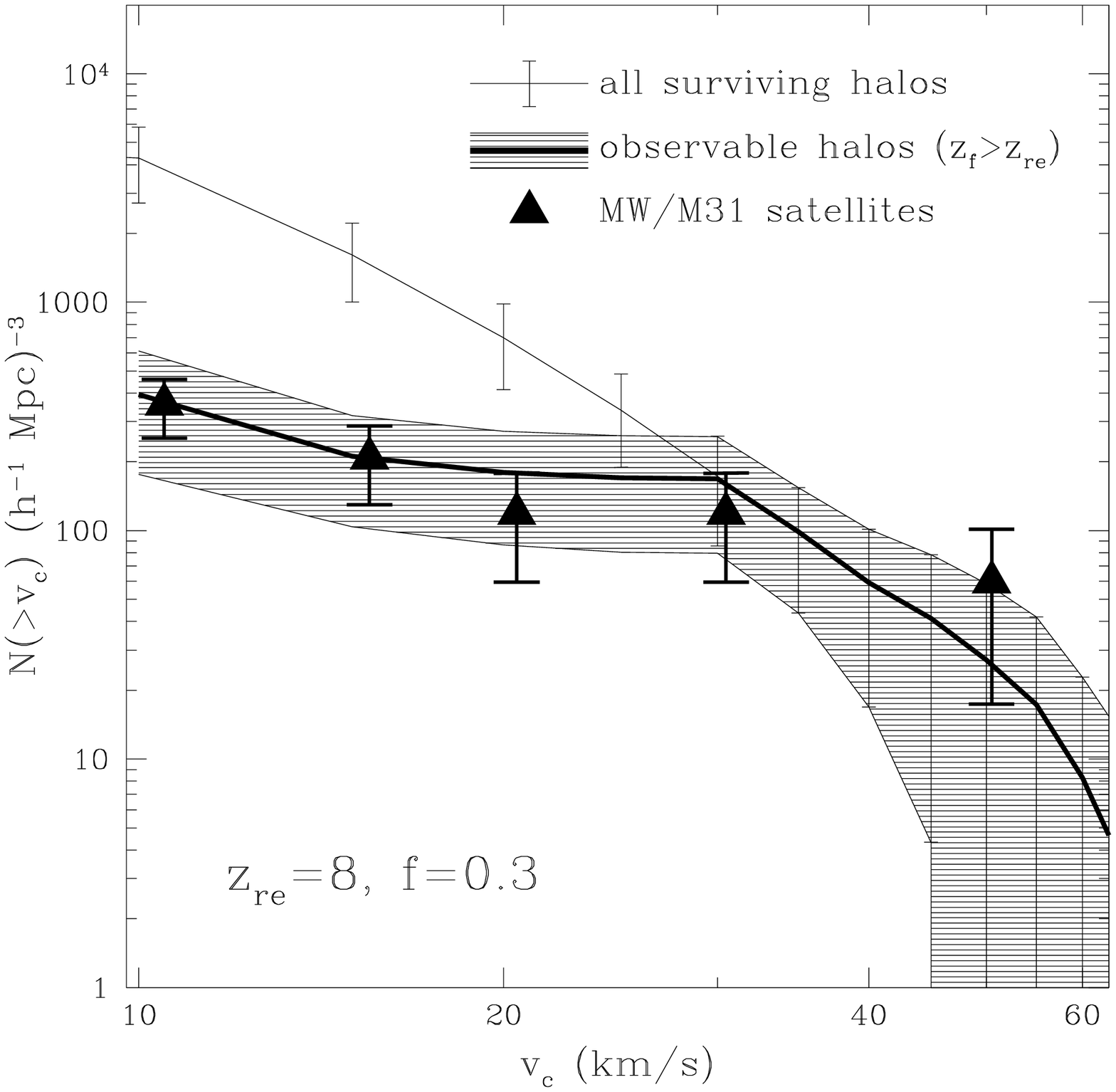}} 
\rput[tl]{0}(0,4.){ 
\begin{minipage}{8.9cm} 
  \small\parindent=4.5mm {\sc Fig.}~2.--- Cumulative velocity function
  of  all dark  matter subhalos  surviving  at $z=0$  ({\em thin solid
  line}) and ``observable''  halos  ($z_{\rm f} > z_{\rm   re}$) ({\em
  thick solid line with shading}),  for the specific choice of $z_{\rm
  re}=8$ and $f=0.3$.  The  velocity function represents  the average
  over    300  merger  histories    for    halos   of mass     $M_{\rm
  vir}(z=0)=1.1\times 10^{12}h^{-1}{\ }{\rm M_{\odot}}$. The errorbars
  and shading  show   the dispersion  measured from different   merger
  histories.  The   observed velocity  function of  satellite galaxies
  around the Milky Way and M31 is shown by triangles.

\end{minipage} 
 } 
\endpspicture} 

\noindent factor of 3 --- raises
the predicted  number  of galaxies by  a factor  of 6.   Reducing  $f$
(increasing  $M/L$) by a  factor of 7 raises  the  predicted number of
satellites by a  factor of 10, and  accounts for $\sim  98 \%$ of  the
dark subhalos.  Large  area, deep imaging  surveys may soon be able to
reveal faint dwarf satellites that lie below current detection limits.

Dark halo satellites  may also  be  detectable by their  gravitational
influence.  For  example,    it   may  be possible to     detect  halo
substructure via gravitational lensing.  Mao \& Schneider (1998) point
out that at least  in the case  of the quadruply  imaged QSO B 1422  +
231, substructure may be  needed in order  to account for the observed
flux ratios.     It is also  plausible   that a large  number  of dark
satellites  could have destructive effects   on disk galaxies (Toth \&
Ostriker  1992; Ibata \& Razoumov 1998;  Weinberg 1998).  Moore et al.
(1999) used a simple calculation based on the impulse approximation to
estimate the amount of heating that the subclumps in their simulations
would produce on  a stellar disk embedded at   the halo center.   They
concluded that this type  of heating is  not problematic  for galaxies
like the Milky  Way, but that the  presence of galaxies with  no thick
disk, such as NGC 4244, may be problematic.  Our results indicate that
the  variation in halo  formation histories is substantial, suggesting
that thin  disks could  occur in recently  formed  systems with little
time   for  significant heating  to    have occurred,  but  a detailed
investigation of this subject is certainly warranted.

It  is tempting to  identify  some  of these   subhalos with the  High
Velocity Clouds (HVCs) (Braun \& Burton 1999; Blitz et  al. 1998).  In
our  scenario, it  would  be possible to   account for  some of  these
objects provided  we modify the model  of \S2.2.  We  currently assume
that there  is no accretion after  reionization, and  that most of the
gas that accretes before $\zre$ is  converted into stars.  However, it
may be that accretion starts again at low redshift, after the level of
the UV background  drops (e.g., Babul  \& Rees 1992; Kepner,  Babul \&
Spergel   1997), and  these  late-accreting  systems might  form stars
inefficiently and retain  their gas as HI.   If this is the case, HVCs
may be associated  with subhalos that are  accreted at late times.  We
find that, on average, $\sim 60\%$ of surviving subhalos fall into the
host halo after $z=1$, and $\sim 40 \%$ fall in after $z=0.5$.

Another prediction   is  that   there should be   a    diffuse stellar
distribution in the Milky Way  halo associated with the disruption  of
many  galactic satellites  (see Figure  1).   If  we  assume that  the
destroyed   subhalos  had the same    stellar content as the surviving
halos, then  we can  estimate  the radial   density  profile of   this
component by  placing the   stars from each   disrupted halo  at   the
apocenter of  its orbit, where  they would spend  most of  their time.
This    calculation  yields   a density   profile  $\rho_*(r)  \propto
r^{-\alpha}$, with $\alpha = 2.5 \pm 0.3$, extending from $r \simeq 10
- 100 \hkpc$,  which is  roughly  consistent with the distribution  of
known   stellar halo populations  such  as  RR Lyrae variables  (e.g.,
Wetterer  \& McGraw 1996).  The normalization  of  the profile is more
uncertain, but for the parameters  $\zre = 8$  and $f = 0.3$, and with
the assumption that each halo with $z_f > \zre$ has a mass in stars of
$M_{*}   = f (\Omega_b/\Omega_0)  \epsilon_* M_{a}$,  we find that the
stellar mass of the disrupted component is $M_*
\sim 5 \times 10^{8} \hMsun$.  This diffuse distribution could make up
a   large   fraction of   the  stellar   halo,  perhaps   all  of  it.
Observationally, it  may   be difficult   to distinguish a   disrupted
population from  a stellar  halo formed  by  other means, but  perhaps
phase space substructure   may  provide  a  useful diagnostic   (e.g.,
Johnston 1998; Helmi et al. 1999).

This  disrupted   population would   not  be  present in  models  with
suppressed small  scale power or warm  dark matter.  However, it would
be   expected in the self-interacting   dark matter scenario.  In this
case,  the   distribution would probably   extend  to a  larger radius
because  dark matter  interactions would  disrupt  the satellite halos
further out.

There are  other problems   facing the CDM   hypothesis, such  as  the
possible   disagreement between the predicted   inner  slopes of  halo
profiles and the rotation curves  of dwarf and  LSB galaxies (Moore et
al. 1999; Kravtsov  et al. 1998; Flores  \& Primack 1994; Moore 1994).
The mechanism proposed  here does not  solve this problem, though more
complicated effects of gas  dynamics and star  formation might  do so.
We have shown that one of the problems facing CDM can be resolved by a
simple  gas dynamical mechanism.  If  this solution is  the right one,
then  the  dark matter structure of   the Milky  Way  halo resembles a
scaled-down version  of a  typical galaxy   cluster, but  most of  the
low-mass Milky Way subhalos formed too late to  accrete gas and become
observable dwarf galaxies.

\acknowledgements 
We thank Andrew      Gould and  Jordi  Miralda-Escud\'e   for   useful
discussions.  This  work was supported   in  part by NASA  LTSA  grant
NAG5-3525 and NSF grant AST-9802568.  Support for A.V.K.  was provided
by NASA through Hubble Fellowship grant HF-01121.01-99A from the Space
Telescope Science Institute, which  is operated by the  Association of
Universities for   Research in Astronomy,  Inc.,  under  NASA contract
NAS5-26555.

\end{document}